\documentclass[aip,11pt,jcp]{revtex4-2}

\makeatletter
\def\@seccntformat#1{\protect\makebox[1cm][l]{\csname the#1\endcsname}}
\makeatother

\usepackage{jpcrd2}

\usepackage{graphicx,epstopdf}
\usepackage{dcolumn}
\usepackage{bm}
\usepackage[utf8]{inputenc}
\usepackage[T1]{fontenc}
\usepackage{mathptmx}
\usepackage{amsmath,amssymb}
\usepackage{setspace}
\usepackage{gensymb}
\usepackage{siunitx}
\usepackage{indentfirst}

\usepackage{color}
\usepackage[colorlinks=true,
linkcolor=blue,
urlcolor=blue,
citecolor=blue,
anchorcolor=blue]
{hyperref}

\newcolumntype{d}[1]{D{.}{.}{#1}}

\begin{document}
\newcommand{\etal}{{\it et al.} }

\title{Recommended Second Virial Coefficients for Nitrogen and Oxygen}

\author{Robert Hellmann}
\affiliation{Institut f\"ur Thermodynamik, Helmut-Schmidt-Universit\"at / Universit\"at der Bundeswehr Hamburg, 22043 Hamburg, Germany, robert.hellmann@hsu-hh.de}

\author{Giovanni Garberoglio}
\affiliation{European Centre for Theoretical Studies in Nuclear Physics and Related Areas (FBK-ECT*), Trento, I-38123, Italy, garberoglio@ectstar.eu}

\author{Allan H. Harvey}
\affiliation{Applied Chemicals and Materials Division,
National Institute of Standards and Technology, Boulder, Colorado, 80305, U.S.A., allan.harvey@nist.gov}

\date{\today}

\newpage
\begin{abstract}
We provide recommended values for the second virial coefficient, $B(T)$, and its uncertainty, for molecular nitrogen and oxygen.
The temperature range covered is 20--3000\;K for nitrogen and 20--2000\;K for oxygen. 
The recommendations are based on tuning previously published state-of-the-art \textit{ab initio} pair potentials so that the $B(T)$ calculated from the potentials match selected high-accuracy experimental data; for nitrogen the tuning utilizes values of $B$ derived from literature density data with greatly reduced uncertainty by analyzing the data with the aid of \textit{ab initio} calculated higher virial coefficients.
Quantum effects on $B$ are fully included with the path-integral Monte Carlo method.
The resulting $B(T)$ have uncertainties similar to those of the best experimental data, but cover a much wider temperature range.

\end{abstract}

\maketitle

\tableofcontents



\newpage

\section{Introduction}\label{sec:intro} 

The vast majority of the Earth's atmosphere consists of molecular nitrogen (N$_2$) and oxygen (O$_2$).
Knowledge of their thermophysical properties is therefore essential for understanding the properties of the mixture known as ``air,'' whose properties are important in many industrial contexts and also in atmospheric science.

In metrology, several contexts require accurate knowledge of air properties.
The density of air is needed for buoyancy corrections for precision weighing.\cite{Davis_1992,Picard_2008}
The thermodynamic properties of dry air are an important component of humidity metrology.\cite{Hyland_1983}
Accurate description of sound propagation in air depends on knowledge of its thermodynamics.\cite{Gavioso_2025}
Both the density and refractivity of air are needed for laser-based length metrology.\cite{Ciddor_1996,Egan_2026,Bartl_2020}
In all of these applications, the degree to which air deviates from ideal-gas behavior is usually described with correlations developed over 40 years ago. The present work is a step toward updating these recommendations based on newer developments in theory and experiment.

Nitrogen and oxygen are also important as pure gases.
One application is flow metering (for example in semiconductor manufacturing), where meters are often calibrated with nitrogen and accurate measurement of other gases (such as oxygen) requires knowledge of how the properties of the other gas differ from those of nitrogen.\cite{Wright_2012}
Nitrogen is also one of the gases applied for pressure standards that employ precision refractometry to measure the number density of a gas, from which its pressure may be derived if the thermodynamic behavior is known.\cite{Jousten_2017,Yang_2021,Ricker_2025}
This technology is the focus of the Joint Research Project ``MQB-Pascal'' under the European Metrology Research Programme,\cite{MQB_web} which provided much of the impetus (and funding) for the work reported here.

In the applications above, the gas is at a relatively low pressure and its deviation from ideality can be accurately described by the virial expansion. The virial expansion provides a series of rigorous corrections to ideal-gas behavior for the pressure $p$ as a function of absolute temperature $T$ and molar density $\rho$:
\begin{equation}
  \frac{p}{\rho R T} = 1 + B(T) \rho + C(T) \rho^2 + D(T) \rho^3 + \cdots,
  \label{eq:virial}
\end{equation}
where $R$ is the molar gas constant.
The lowest-order correction, $B(T)$, is the second virial coefficient; it is a function of the interaction between two molecules. The third virial coefficient, $C(T)$, depends on interactions between two molecules and among three molecules, $D(T)$ depends on interactions among two, three, and four molecules, etc.
For applications near typical atmospheric pressure ($\sim 0.1$\;MPa), it is often sufficient to truncate the virial expansion after the $B(T)$ term.
Equation (\ref{eq:virial}) is also valid for mixtures at a fixed composition, and each virial coefficient is rigorously given as a function of temperature-dependent coefficients for subsystems in the mixture. For example, for the second virial coefficient
\begin{equation} \label{eq:Bmix}
  B_\mathrm{mix}(T) = \sum_{i,j} y_i y_j B_{ij}(T) ,
\end{equation} 
where $y_i$ is the mole fraction of component $i$.
The terms in Eq. (\ref{eq:Bmix}) where $i = j$ are simply the pure-component virial coefficients. The $B_{ij}$ with unlike subscripts are ``cross'' coefficients that correspond to the interaction between one molecule of species $i$ and one molecule of species $j$.

It is clear from Eqs. (\ref{eq:virial}) and (\ref{eq:Bmix}) that accurate thermodynamic description of air requires $B(T)$ for both pure N$_2$ and pure O$_2$.
It also requires the cross second virial coefficient between N$_2$ and O$_2$; that will be a topic of future work.
The goal of this work is to provide recommended values of $B(T)$ and its uncertainty for both N$_2$ and O$_2$.

For pure fluids, $B(T)$ can be obtained from measurements of density as a function of pressure and temperature.  If the data are sufficiently accurate and appropriate attention is paid to details of the analysis of $\rho(p)$ along isotherms, accurate values of $B$ can be derived.
However, this approach requires extensive experiments in order to obtain a single value of $B$, resulting in a small number of data points that inevitably have some scatter and cover a limited temperature range.

An alternative approach employs \textit{ab initio} quantum chemistry to develop a potential-energy surface (PES) between the molecules.  Once such a surface exists, $B(T)$ can be calculated rigorously at any temperature. 
This approach has been very successful for the noble gases,\cite{Garberoglio_2023,Harvey_2025} with uncertainties obtained from theory 
much smaller than those of the best experiments for helium,\cite{Czachorowski_2020} smaller than those of almost all experimental data for neon,\cite{Hellmann_2021} and modestly exceeding those from the best experiments for argon.\cite{Lang_2024}
For diatomic molecules such as N$_2$ and O$_2$, \textit{ab initio} calculations are more difficult and uncertainties in $B$ are typically somewhat larger than those of the best experiments.  Since the experiments are typically only at a few temperatures, a fruitful approach to cover a wide temperature range is to use the experimental data to ``tune'' a pair potential by slightly adjusting a parameter in the representation of the \textit{ab initio} PES to produce a function $B(T)$ that matches the selected experiments and covers the entire temperature range of interest.

For nitrogen, this approach was followed by Hellmann in 2013,\cite{Hellmann_2013} who calculated the PES between two N$_2$ molecules (considered as rigid rotors) at the CCSDT(Q) (coupled cluster with single, double, triple, and perturbative quadruple excitations) level of theory, with large basis sets extrapolated to the complete-basis-set limit, and including relativistic as well as core--core and core--valence correlation effects.
A small adjustment to the PES was made in order to better reproduce the best $B(T)$ data, particularly those reported by Nowak \textit{et al.}\cite{Nowak_1997} based on their state-of-the-art density measurements at pressures up to 12 MPa and temperatures 98--340~K.

There is now better information available to refine the tuning of the nitrogen PES.
Values of $B(T)$ for N$_2$ with small uncertainties were reported by Egan and Yang\cite{Egan_2023} based on refractivity measurements from 293--433~K.
Accurate density isotherms were reported by McLinden and L\"{o}sch-Will.\cite{McLinden_2007}
Perhaps most important, it is now possible to reanalyze isotherms from previous $\rho(p)$ experiments\cite{Nowak_1997,McLinden_2007} using fourth and higher-order virial coefficients calculated from the pair potential and a new \emph{ab initio} three-body PES developed also within the MQB-Pascal project by one of us (RH), which is available upon reasonable request.  Using these physically reasonable values of $D$, $E$, etc.\ to constrain the analysis of the density data produces $B$ and $C$ on each isotherm with significantly smaller uncertainty than previous analysis in which the higher coefficients were free parameters or assumed to be zero. 
Such a procedure has recently been applied to accurate $\rho(p)$ experimental data for neon.\cite{Hellmann_2026}

Similar considerations apply for oxygen, where Hellmann developed a state-of-the art PES (actually three PES, corresponding to the singlet, triplet, and quintet interaction surfaces of the triplet O$_2$ molecule) and used it to compute $B(T)$.\cite{Hellmann_2023}
In this case, there were no experimental data accurate enough to tune the potential.
Since that work, Egan has reported $B(T)$ for oxygen from 293--373~K based on precision refractometry.\cite{Egan_2026}  These data can be used to refine the PES and therefore $B(T)$.

In Sec.~\ref{sec:potentials}, we describe the reanalysis of experimental data and the tuning of the pair potentials for both molecules to these and other recent data.
In Sec.~\ref{sec:calc}, we describe the computation of $B(T)$ from the potentials, rigorously accounting for quantum effects with the path-integral approach.
The results for $B(T)$ are presented and compared to select experimental values in Sec.~\ref{sec:results}, while Sec.~\ref{sec:unc} describes our estimation of their uncertainty.
In Sec.~\ref{sec:recc}, we present recommended functions for $B(T)$ and its expanded uncertainty for both substances.
Finally, Sec.~\ref{sec:concl} will summarize the work and discuss future use of these results in constructing important functions such as $B(T)$ for both dry and humid air.


\section{Tuning of Pair Potentials}\label{sec:potentials}

\subsection{Nitrogen}\label{sec:potentials_N2}

The main pillar for our retuning of the PES between two nitrogen molecules is a careful reanalysis of most of the $\rho(p)$ isotherms of Nowak \textit{et al.},\cite{Nowak_1997} which were measured with a dual-sinker densimeter and a piston gauge and are widely considered as the most accurate and comprehensive density data available for this fluid. In addition, we reanalyzed the more limited but equally accurate $\rho(p)$ isotherms of McLinden and L\"{o}sch-Will\cite{McLinden_2007} obtained with the same technique as those of Nowak \textit{et al.} There is a third data set measured with this approach, from Duschek \textit{et al.}\cite{Duschek_1988} However, we deemed this set not worth reanalyzing because it was measured with the same apparatus as Nowak \textit{et al.} but in a much narrower temperature and density range.

We calculated the third to sixth virial coefficients, i.e., $C$, $D$, $E$, and $F$, at the temperatures at which Nowak \textit{et al.}\ and McLinden and L\"{o}sch-Will performed their $\rho(p)$ measurements, using the quadratic Feynman--Hibbs\cite{Feynman_1965} (QFH) modification of a version of the pair PES tuned to the data of Egan and Yang.\cite{Egan_2023} Nonadditive three-body interactions were accounted for by the aforementioned \emph{ab initio} three-body potential. The computational details, also for the semiclassical calculations of $B$ discussed later, are similar to those reported by Hellmann for carbon dioxide,\cite{Hellmann_2014,Hellmann_2017} so we omit them here.

We reanalyzed 20 isotherms of Nowak \textit{et al.}\ from 130--340\;K. We did not reanalyze their isotherms between 98\;K and 130\;K because we observed a strong increase in the uncertainties of the derived $B$ and $C$ with decreasing temperature for isotherms below 150\;K. Derived values below 130\;K would have uncertainties too high for the intended tuning of the pair potential. Each considered isotherm was fitted as $p(\rho)$ by a sixth-order virial expansion in which $D$, $E$, and $F$ were constrained to the respective semiclassically calculated values, while $B$ and $C$ and initially also the zero-density limit of $p/(\rho RT)$ were fitted. The latter should be unity (the universal ``first'' virial coefficient), but the actual value differs due to small systematic density and pressure measurement errors as well as due to the actual (unknown) molar mass of the nitrogen sample differing from the value assumed when converting the measured mass density to the molar density. Assuming a molar mass of 28.014\;g/mol, we obtained $\lim_{\rho\rightarrow0}p/(\rho RT)$ ranging from 0.9999258 to 0.9999943, with an average value of 0.9999569. When plotting $\lim_{\rho\rightarrow0}p/(\rho RT)$ \emph{vs.}\ $T$, despite significant scatter, we saw a clear trend towards higher values at higher temperatures. 
Since temperature does not affect the molar mass and because it is standard practice to correct the sinker volumes for thermal expansion, we saw no reason 
to allow for anything but a constant, temperature-independent value of $\lim_{\rho\rightarrow0}p/(\rho RT)$ in our analysis, for which we used the average value. 

During the analysis, we observed that sometimes the measured state points at the lowest densities and pressures deviated substantially from any reasonable virial expansion fits, which prompted us to remove these points before continuing the analysis. These outliers are the lowest density at 133\;K, 155\;K, 162\;K, 180\;K, 200\;K, 280\;K, 293.15\;K, 320\;K, and 340 K, the two lowest densities at 140\;K and 150\;K, and the three lowest densities at 130\;K.

In the temperature range 280--340\;K, all state points up to the highest experimentally investigated pressure of 12\;MPa, except for the mentioned low-pressure outliers, were included in the virial fits because the resulting values of $B$ and $C$ for a given isotherm clearly converged to stable constant values when including higher and higher pressures in the fits. For the 260\;K isotherm, this convergence was only observed up to pressures of 11\;MPa, with the inclusion of the two 12\;MPa state points resulting in changes to $B$ and $C$ that are clearly not within the scatter of the converged values when fitting with lower maximum pressures. We attribute this to virial coefficients beyond $F$ starting to contribute significantly above 11\;MPa at this temperature. The lower the temperature, the more we had to reduce the maximum pressure included in the fit (to 10\;MPa at 240\;K, 8\;MPa at 220\;K, 7\;MPa at 200\;K and 190\;K, 6\;MPa at 180\;K, 5.5\;MPa at 170\;K, 5\;MPa at 162\;K and 155\;K, 4.8\;MPa at 150\;K, 4.3\;MPa at 145\;K and 140\;K, 4\;MPa at 136\;K and 133\;K, and 3.7\;MPa at 130\;K). At temperatures below 170\;K, the plateaus in $B$ and $C$ that we observed at higher temperatures when plotting these coefficients against the highest pressure included in the virial expansion fits became more and more unstable. However, due to the steep $\rho^6$ dependence of the unaccounted seventh virial coefficient, it was always relatively clear from the plots what the highest pressure for a given isotherm should be such that the unaccounted virial coefficients beyond the sixth do not significantly distort the values for $B$ and $C$.

The uncertainties of the virial coefficients $B$ and $C$ extracted from the data of Nowak \textit{et al.} by this approach were estimated by adding three separate contributions in quadrature. One contribution was simply the difference obtained by performing the virial fits with and without constraining $\lim_{\rho\rightarrow0}p/(\rho RT)$ as described above. While we argued that constraining to a single temperature-independent value should be the right choice, we cannot confidently rule out that unconstrained fits result in more accurate $B$ and $C$. What we saw, however, is that the unconstrained values exhibit a much stronger scatter relative to a smooth baseline such as the values obtained from the potentials. 
Another uncertainty contribution is due to the uncertainties in the calculated virial coefficients $D$, $E$, and $F$ used for the analysis. For simplicity, we only considered the uncertainty of the most important of these three coefficients, $D$, which we estimated to be 2000\;cm$^9$\,mol$^{-3}$. This value is realistic for temperatures below $\sim200$\;K and becomes more and more conservative toward 340\;K. The uncertainty contributions for $B$ and $C$ due to the uncertainty in $D$ were then obtained as the differences between the values obtained with and without adding 2000\;cm$^9$\,mol$^{-3}$ to $D$. 
The third uncertainty contribution is related to how stable the $B$ and $C$ values are within the abovementioned plateaus. It is thus mostly related to the statistical uncertainty of the underlying experimental data and was estimated for both $B$ and $C$ as the difference between the largest and smallest value within the (more or less) stable plateau between too small and too high maximum pressures (where the highest maximum pressure in the plateau is the one used to obtain the final $B$ and $C$ values). Assigning a reasonable lowest maximum pressure that is still part of the plateau is admittedly quite subjective for many of the isotherms. Due to the strong negative correlation that $B$ and $C$ have as fit coefficients (a too large $B$ value can be partly compensated by a too small $C$ value and vice versa), the pressure range assigned to the plateau on each isotherm was the same for $B$ and $C$. We consider the final uncertainties as expanded uncertainties with coverage factor $k=2$, corresponding approximately to a 95\,\% confidence level.

McLinden and L\"{o}sch-Will\cite{McLinden_2007} measured five isotherms with nominal temperatures from 293.15\;K to 480\;K at pressures up to 6\;MPa with a combination of a dual-sinker densimeter and a piston gauge. They also performed measurements up to much higher pressures but using far less accurate pressure transducers for the pressure measurement; we did not consider these data for our analysis. The isotherms are actually not perfect isotherms as the temperatures vary by a few millikelvin between individual state points. While this is an essentially negligible error source, we still corrected the measured pressures by the amount expected to result from changing the temperature isochorically to the average temperature using the ideal-gas law. With 74 state points distributed over 12 different nominal pressures, the 293.15\;K isotherm is the most comprehensive, but the five points at the lowest nominal pressure of 0.1\;MPa are outliers and therefore had to be discarded. The other four isotherms, which have nominal temperatures of 340\;K, 400\;K, 440\;K, and 480\;K, comprise only between 22 and 24 state points distributed over five nominal pressures (0.5\;MPa, 1\;MPa, 2\;MPa, 4\;MPa, and 6\;MPa), and the 0.5\;MPa state points at 400\;K and 440\;K had to be discarded as outliers. The virial expansion fits of the 293.15\;K, 340\;K, and 400\;K isotherms were performed in a similar manner as for those of Nowak \textit{et al.}\cite{Nowak_1997} The only difference was that we constrained $\lim_{\rho\rightarrow0}p/(\rho RT)$ to unity because it could not be reliably determined for any of these isotherms, not even for the 293.15\;K isotherm, which has significant scatter in the fit residuals below 2\;MPa. For the 440\;K and 480\;K isotherms, in addition to constraining $\lim_{\rho\rightarrow0}p/(\rho RT)$ to unity, we constrained $C$ (not only $D$, $E$, and $F$) to the semiclassically calculated values and extracted only $B$. At these high temperatures, even the densities at 6\;MPa are too small to extract reasonably accurate values for $B$ and $C$ together.

The uncertainty estimation for $B$ and $C$ extracted from the measurements of McLinden and L\"{o}sch-Will was performed in a similar manner as for the values extracted from the Nowak \textit{et al.}\ data. However, for the four isotherms above 293.15\;K, the small number of nominal pressures prevented identification of plateaus in $B$ and $C$ when plotting them against the maximum pressure included in the fits. For these isotherms, the uncertainty contributions due to the statistical uncertainties of the underlying experimental data were estimated as the differences in the $B$ and $C$ values resulting from fits up to nominal pressures of 4\;MPa and 6\;MPa. For the $B$ values at 440\;K and 480\;K, the uncertainty contributions due to the uncertainties of the higher virial coefficients used to constrain the fits were determined by varying $C$ by a conservative estimate for its uncertainty of 15\;cm$^6$\,mol$^{-2}$ at these temperatures.

The $B$ and $C$ values derived in this work from experimental $\rho(p)$ isotherms of Nowak \textit{et al.}\cite{Nowak_1997} and McLinden and L\"{o}sch-Will\cite{McLinden_2007} are provided together with their expanded ($k=2$) uncertainties and the semiclassically calculated values of $C$, $D$, $E$, and $F$ in Table~\ref{tab:Virial_exp}.
\begin{table}
\footnotesize
\caption{\label{tab:Virial_exp} Values for the second virial coefficient $B$ and its estimated expanded ($k=2$) uncertainty $U(B)$ and for the third virial coefficient $C$ and its estimated expanded ($k=2$) uncertainty $U(C)$ obtained by a reanalysis of $\rho(p)$ isotherms of Nowak \textit{et al.}\cite{Nowak_1997} and of McLinden and L\"{o}sch-Will\cite{McLinden_2007} as well as semiclassically calculated values of $C$ and their deviations from the experimental ones and semiclassically calculated values of the higher virial coefficients $D$, $E$, and $F$.} 
\begin{ruledtabular}
\begin{tabular}{d{3.3}d{3.3}d{1.3}d{4.0}d{3.0}d{4.0}d{3.0}d{1.3}d{3.2}d{3.1}}
\multicolumn{1}{c}{$T$} &
\multicolumn{1}{c}{$B_\mathrm{exp}$} &
\multicolumn{1}{c}{$U\left(B_\mathrm{exp}\right)$} &
\multicolumn{1}{c}{$C_\mathrm{exp}$} &
\multicolumn{1}{c}{$U\left(C_\mathrm{exp}\right)$} &
\multicolumn{1}{c}{$C_\mathrm{calc}$} &
\multicolumn{1}{c}{$C_\mathrm{calc}-C_\mathrm{exp}$} &
\multicolumn{1}{c}{$10^{-4}\times D_\mathrm{calc}$} &
\multicolumn{1}{c}{$10^{-5}\times E_\mathrm{calc}$} &
\multicolumn{1}{c}{$10^{-7}\times F_\mathrm{calc}$}
\\
\cline{2-3}\cline{4-7}
\multicolumn{1}{c}{(K)} &
\multicolumn{2}{c}{(cm$^3$\,mol$^{-1}$)} &
\multicolumn{4}{c}{(cm$^6$\,mol$^{-2}$)} &
\multicolumn{1}{c}{(cm$^9$\,mol$^{-3}$)} &
\multicolumn{1}{c}{(cm$^{12}$\,mol$^{-4}$)} &
\multicolumn{1}{c}{(cm$^{15}$\,mol$^{-5}$)}
\\[0.5ex]
\hline
\\[-1.5ex]
\multicolumn{10}{c}{Nowak \emph{et~al.}\cite{Nowak_1997}} \\[1ex]
130.000 & -96.341 & 0.068 & 2941 &  35 & 2893 & -48 & 6.002 &  -7.26 & -18.1 \\
133.000 & -92.004 & 0.057 & 2882 &  24 & 2839 & -43 & 5.498 &  -9.61 & -17.1 \\
136.000 & -87.921 & 0.048 & 2822 &  25 & 2782 & -40 & 5.005 & -11.62 & -16.0 \\
140.000 & -82.837 & 0.045 & 2744 &  22 & 2704 & -40 & 4.391 & -13.58 & -14.4 \\
145.000 & -76.983 & 0.058 & 2645 &  32 & 2607 & -38 & 3.708 & -14.95 & -12.0 \\
150.000 & -71.626 & 0.032 & 2550 &  17 & 2512 & -38 & 3.129 & -15.29 &  -9.5 \\
155.000 & -66.697 & 0.027 & 2457 &  16 & 2421 & -36 & 2.651 & -14.88 &  -7.0 \\
162.000 & -60.433 & 0.030 & 2334 &  16 & 2301 & -33 & 2.133 & -13.46 &  -4.0 \\
170.000 & -54.059 & 0.022 & 2208 &  14 & 2177 & -31 & 1.718 & -11.19 &  -1.2 \\
180.000 & -47.063 & 0.022 & 2068 &  14 & 2041 & -27 & 1.402 &  -8.06 &   1.3 \\
190.000 & -40.971 & 0.023 & 1949 &  15 & 1924 & -25 & 1.245 &  -5.04 &   2.9 \\
200.000 & -35.605 & 0.020 & 1845 &  13 & 1824 & -21 & 1.192 &  -2.35 &   3.9 \\
220.000 & -26.629 & 0.022 & 1682 &  14 & 1664 & -18 & 1.267 &   1.87 &   4.6 \\
240.000 & -19.434 & 0.023 & 1564 &  14 & 1547 & -17 & 1.450 &   4.75 &   4.7 \\
260.000 & -13.531 & 0.022 & 1474 &  14 & 1460 & -14 & 1.660 &   6.63 &   4.4 \\
280.000 &  -8.616 & 0.025 & 1407 &  15 & 1394 & -13 & 1.863 &   7.82 &   4.1 \\
293.150 &  -5.806 & 0.039 & 1370 &  16 & 1359 & -11 & 1.985 &   8.35 &   3.8 \\
300.000 &  -4.466 & 0.028 & 1354 &  14 & 1343 & -11 & 2.044 &   8.56 &   3.7 \\
320.000 &  -0.919 & 0.026 & 1313 &  14 & 1304 &  -9 & 2.201 &   8.99 &   3.4 \\
340.000 &   2.135 & 0.022 & 1281 &  13 & 1273 &  -8 & 2.333 &   9.21 &   3.1 \\[1ex]
\multicolumn{10}{c}{McLinden and L\"{o}sch-Will\cite{McLinden_2007}} \\[1ex]
293.139 &  -5.827 & 0.016 & 1371 &   9 & 1359 & -12 & 1.985 &   8.35 &  3.8 \\
339.976 &   2.139 & 0.033 & 1276 &  15 & 1273 &  -3 & 2.333 &   9.21 &  3.1 \\
399.973 &   9.186 & 0.029 & 1213 &  19 & 1210 &  -3 & 2.604 &   9.20 &  2.5 \\
439.979 &  12.597 & 0.063 &      &     & 1183 &     & 2.706 &   8.93 &  2.2 \\
480.003 &  15.337 & 0.056 &      &     & 1163 &     & 2.767 &   8.58 &  1.9
\end{tabular}
\end{ruledtabular}
\end{table}
The excellent agreement between the $B$ and $C$ values derived from both experimental studies at the two common nominal temperatures 293.15\;K and 340\;K strongly validates the correctness of the underlying density data. While the calculated $C$ values (which, except for the highest two temperatures of McLinden and L\"{o}sch-Will, were not used in the reanalysis of the experimental data) are systematically smaller than the experimental ones, the agreement is still very satisfactory. This provides confidence in the calculated higher virial coefficients.

Our tuning strategy for the pair PES follows that of the paper of Hellmann,\cite{Hellmann_2013} who scaled the most uncertain contribution to the interaction energy, namely, that accounting for the difference between the CCSDT(Q) and CCSD(T) levels of theory, by a constant factor of 0.5, after which he refitted the PES. With the newly derived $B$ values given in Table~\ref{tab:Virial_exp} and the recent $B$ values of Egan and Yang,\cite{Egan_2023} we refined the scaling factor to 0.472.
This tuning was dominated by the values from Table~\ref{tab:Virial_exp}, since their uncertainties are somewhat smaller than those of Ref.~\citenum{Egan_2023}.
The aforementioned preliminary version of the pair PES tuned solely to the Egan and Yang values used a scaling factor of 0.52. 
For the refitting, we found it sufficient to reoptimize only the linear fit parameters. A Fortran~90 routine of the PES is provided in the \textcolor{blue}{Supplementary Material}. The $B$ values obtained with the retuned pair PES are compared with the experimentally derived values used for the tuning and a few other high-quality experimental data in Sec.~\ref{sec:results}.

\subsection{Oxygen}\label{sec:potentials_O2}

When two O$_2$ molecules in their $^3\Sigma_g^-$ ground electronic states are close enough for their orbitals containing the unpaired electrons to overlap, their spins couple such that the molecule pair is either in a quintet state (with a statistical probability of 5/9), a triplet state (probability 3/9), or a singlet state (probability 1/9). In each of these three cases, the two molecules interact through a different pair potential. The second virial coefficient $B$ is obtained as the sum of contributions from each surface with the respective statistical weight,
\begin{equation}\label{Bsum}
B = \frac{5}{9} B_\mathrm{quintet} + \frac{3}{9} B_\mathrm{triplet} + \frac{1}{9} B_\mathrm{singlet},
\end{equation}
where we note that the $B$ quantities on the right-hand side are not experimentally observable.

Hellmann\cite{Hellmann_2023} developed analytical representations of all three potentials based on highly accurate \emph{ab initio} calculations. He did not tune the potentials because the agreement of his calculated $B(T)$ values with the best available experimental data was already satisfactory. However, we now have new experimental $B(T)$ values of Egan from 293--373~K based on precision refractometry.\cite{Egan_2026} They have an expanded ($k=2$) uncertainty of 0.1\;cm$^3$\,mol$^{-1}$, which is significantly smaller than previously achieved for oxygen. Therefore, we retuned the three potentials of Hellmann to match these data as well as possible. 
The employed strategy is different than for nitrogen, but it also has effectively only a single tuning parameter. For each analytical PES, we varied the isotropic part of the $C_8$ dispersion coefficient, i.e., the $C_8$ dispersion parameter for the interaction between the sites in the centers of mass of the two molecules, such that $B_\mathrm{quintet}$, $B_\mathrm{triplet}$, and $B_\mathrm{singlet}$ at 293.15\;K increase by 0.27\;cm$^3$\,mol$^{-1}$. The required parameter changes are quite different for the three surfaces, since the parameters governing the damping of the respective dispersion terms at short separation differ significantly. Therefore, and because we have three unique surfaces to begin with, one might suspect that the resulting changes in $B_\mathrm{quintet}$, $B_\mathrm{triplet}$, and $B_\mathrm{singlet}$ will become more and more dissimilar as temperature increases. However, this is not the case. Even at 373.15\;K, $B_\mathrm{quintet}$, $B_\mathrm{triplet}$, and $B_\mathrm{singlet}$ change by nearly the same amount, 0.20\;cm$^3$\,mol$^{-1}$. This demonstrates that the temperature dependence of the changes in $B$ resulting from the tuning is insensitive to the details of the tuning procedure, justifying the use of different one-parameter tuning procedures not only in this work but also in many previous \emph{ab initio} studies on other gases and gas mixtures.

We note that, at temperatures above roughly 2000~K, the population of the first excited electronic state ($a^1\Delta_g$) of O$_2$ begins to become significant, so that additional interaction surfaces would contribute to $B(T)$. 
We therefore only perform calculations for O$_2$ up to 2000~K, compared to 3000~K for N$_2$.

Fortran~90 routines of the three tuned potentials are provided in the \textcolor{blue}{Supplementary Material}. The $B$ values obtained with these potentials are compared with the values of Egan and some older experimentally derived values in Sec.~\ref{sec:results}.


\section{Calculation of Second Virial Coefficient}\label{sec:calc}
\newcommand{\kB}{k_\mathrm{B}}
\newcommand{\QB}{Q_{1,\mathrm{B}}}
\newcommand{\Qxc}{Q_{1,\mathrm{xc}}}
\newcommand{\QBtwo}{Q_{2,\mathrm{B}}}
\newcommand{\Qxctwo}{Q_{2,\mathrm{xc}}}
\newcommand{\qrot}{q_\mathrm{rot}}
\newcommand{\tr}{\mathrm{tr}}
\newcommand{\ee}{\mathrm{e}}
\newcommand{\dd}{\mathrm{d}}

\newcommand{\mbe}{\mathbf{e}}
\newcommand{\mbr}{\mathbf{r}}
\newcommand{\mbu}{\mathbf{u}}
\newcommand{\mbR}{\mathbf{R}}

\subsection{Single-molecule partition function}

The partition function of a homonuclear diatomic molecule can be written as~\cite{Lang26}
\begin{equation}
  Q_1 = \frac{1}{2} \left( Q_\mathrm{B} + Q_\mathrm{xc} \right),
  \label{eq:Q1}
\end{equation}
where the first term (Boltzmann contribution) assumes {\em distinguishable} nuclei and the second
term depends on the fermionic or bosonic nature of the nuclei. Here we consider two ${}^{14}$N
atoms, which have nuclear spin $S=1$ and are therefore bosons.
However, the temperature dependence of the two terms in Eq.~(\ref{eq:Q1}) is markedly different: the ratio $\Xi(T) = Q_\mathrm{xc} / Q_\mathrm{B}$ decreases very rapidly with increasing
temperature,~\cite{Garberoglio2014,Garberoglio18:water} and in the case of nitrogen $Q_\mathrm{xc}$ is negligible for $T \gtrsim 10$~K,~\cite{Lang26} and hence will be neglected in the following.

Additionally, we will consider a rigid molecular model of both molecules so that $H_1$ becomes
the Hamiltonian of a rigid linear rotor,
\begin{equation}
  H_1 = \frac{\hslash^2}{2 I} \hat{\mathbf{J}}^2,
\end{equation}
where $I = \mu_1 \ell^2$ is the inertia moment of the rigid molecule as a function of the reduced mass
of the two atoms, $\mu_1$, and the bond length $\ell$. For N${}_2$ we used $\mu_1 =
7.001537$~amu (1 amu = $1.660539069 \times 10^{-27}$~kg)\cite{CODATA_2025} and $\ell = 0.11014$~nm, while for O${}_2$ we used $\mu_1 = 7.9974575$~amu and $\ell =
0.12109$~nm.  In this approximation, the eigenfunctions of the single-molecule Hamiltonian are the
spherical harmonics, and the corresponding eigenvalues are $E_J = \hslash^2 J (J+1)/(2 I)$, each
with $2J+1$ degeneracy. Since we consider the nuclei distinguishable, there is no constraint on
the values of $J$. The rotational partition function of a single diatomic molecule in this
approximation is then
\begin{equation}
  q_\mathrm{rot} = (2S+1)^2 \sum_{J=0}^\infty (2J+1) \exp\left[-\frac{\beta \hslash^2}{2I} J (J+1) \right],
\end{equation}
where $S$ is the nuclear spin of one of the two atoms.

\subsection{Second virial coefficient of two rigid rotors}

The second virial coefficient of two molecules is given by~\cite{HCB}
\begin{equation}
  B(T) = - \frac{1}{2{\cal V}} \frac{\Lambda^6}{\qrot^2} \left(
  2 Q_2 - {\cal V}^2
  \right),
  \label{eq:B}
\end{equation}
where $\cal V$ is the volume of the box containing the molecules (with the limit ${\cal V} \to
\infty$ taken at the end of the calculations) and
\begin{equation}
  Q_2 = \frac{1}{2} \left[ \tr\left( \ee^{-\beta H_2} \right)
    + \tr\left(\ee^{-\beta H_2} {\cal P} \right) \right]
  \label{eq:Q2}
\end{equation}
is the partition function of the two molecules. In Eq.~(\ref{eq:Q2}), $H_2$ is the interaction
Hamiltonian, and $\cal P$ is the operator exchanging the two molecules.

As demonstrated in Ref.~\onlinecite{Garberoglio2025}, the intermolecular exchange term in
Eq.~(\ref{eq:Q2}) becomes negligible for D${}_2$ at $T \gtrsim 10$~K; this is {\em a fortiori} true
for N${}_2$ and O${}_2$. We thus omit this term hereafter, as our analysis concerns temperature
ranges where its contribution is insignificant. We are left with
\begin{equation}
  B(T) = -\frac{1}{2 {\cal V}} \left[ \frac{\Lambda^6}{\qrot^2}
    \tr\left(\ee^{-\beta H_2}\right)
    - {\cal V}^2 \right].
  \label{eq:B_rigid}
\end{equation}

For a rigid molecular model, the trace in Eq.~(\ref{eq:B_rigid}) is performed over the
nuclear spin states, the center-of-mass coordinates $\mbr_1$ and $\mbr_2$, and the orientations $\mbu_1$ and
$\mbu_2$ of the two molecules. The calculation of $B(T)$ from Eq.~(\ref{eq:B_rigid}) is conveniently
performed using the path-integral approach to quantum statistical mechanics,~\cite{Feynman_1965} as discussed thoroughly
in the literature.~\cite{Patkowski2008,Garberoglio2012b,Garberoglio2025} In this approach, the
quantum partition functions in Eq.~(\ref{eq:B_rigid}) are evaluated via a classical mapping with $P$
replicas, which is exact for $P \to \infty$. We will denote the coordinates of each replica as
$\mbr_1^{(k)}$, $\mbr_2^{(k)}$, $\mbu_1^{(k)}$ and $\mbu_2^{(k)}$, with $k = 1, 2, \ldots P$. The molecules within
each replica interact via the molecular pair potential $V_2\left(\mbr_1^{(k)}, \mbr_2^{(k)}; \mbu_1^{(k)},
\mbu_2^{(k)}\right)$, while the molecules among different replicas interact with effective quantum
potentials connecting each replica $k$ with the subsequent one $k+1$ using periodic boundary
conditions, that is, replica $P+1$ is identified with replica $1$.

The final expression for $B(T)$ is equivalent to the classical expression for a pair of ring
polymers of $P$ beads and is given by
\begin{equation}
  B(T) = -\frac{\Lambda^3}{2} \int
  \left\langle
  \exp\left(-\beta \overline{V_2}\right) - 1
  \right\rangle
  \dd^3\mbR,
  \label{eq:B_PIMC}
\end{equation}
where we have defined
\begin{equation}
  \overline{V_2} = \frac{1}{P} \sum_{k=1}^P
  V_2\left(\mbr_1^{(k)}, \mbr_2^{(k)}; \mbu_1^{(k)}, \mbu_2^{(k)}\right),
  \label{eq:Vbar}
\end{equation}
where the brackets indicate an average over all the translational and rotational internal coordinates of the two ring polymers using probability distributions specified by the quantum-to-classical mapping of the path-integral approach.~\cite{Marx99,Garberoglio08} 
In Eq.~(\ref{eq:B_PIMC}) we have defined $\mbR = \mbr_2^{(0)} - \mbr_1^{(0)}$.

Equations~(\ref{eq:B_PIMC}) and (\ref{eq:Vbar}) suggest the possibility of approximating $B(T)$
using suitably chosen effective potentials. In the high-temperature regime, Feynman and Hibbs\cite{Feynman_1965}
showed that one can approximate $\left\langle \ee^{-\beta \overline{V_2}} \right\rangle$ with
$\left \langle \ee^{-\beta V_\mathrm{QFH}}\right\rangle_{\mbu_1,\mbu_2}$, where
\begin{equation}
  V_\mathrm{QFH} = V_2 + \frac{\hslash^2 \beta}{24} \left(
  \frac{1}{\mu_2} \sum_{j=1}^3 \frac{\partial^2 V_2}{\partial \mbr_j^2} +
  \frac{1}{I} \sum_{i=1,2} \sum_{j=1,2} \frac{\partial^2 V_2}{\partial \theta_{i,j}^2}
  \right),
  \label{eq:VQFH}  
\end{equation}
$\mu_2 = m/2$ is the reduced mass of the pair of molecules,
$\theta_{i,j}$ is the rotation angle of
molecule $i$ around the axis $\mbe_{i,j}$ ($j=1,2$), where each $\mbe_{i,j}$ lies in the plane orthogonal to the direction
$\mbu_i$ ($\mbe_{i,j} \cdot \mbu_i = 0$) and the two $\mbe_{i,j}$ corresponding to molecule $i$ are mutually orthogonal
($\mbe_{i,1} \cdot \mbe_{i,2} = 0$).
Finally, the average $\langle \cdots \rangle_{\mbu_1, \mbu_2}$ is performed over uniformly distributed
orientations of the two molecules.
Note that in the limit $T \to \infty$ one has $V_\mathrm{QFH} \to V_2$ and the classical expression for the second virial coefficient is recovered.
This approximation has been used in deriving the semiclassical virial coefficients in
Secs.~\ref{sec:potentials_N2} and \ref{sec:results}.

\subsection{Calculation details}

Although the path-integral expression for $B(T)$ given in Eq.~(\ref{eq:B_PIMC}) is exact in the $P \to
\infty$ limit, in practice convergence is reached for large enough $P$. 
The value of $P$ needed depends on the accuracy required. In this work, we aimed at an accuracy on the
order of 10~ppm (1 ppm = $10^{-6}$) or less for the widest possible temperature range, which we were able to reach by
setting $P = \mathrm{nint}\left( 20 + 6000~\mathrm{K}/T\right)$ for both N${}_2$ and O${}_2$, where
$\mathrm{nint}(x)$ denotes the integer closest to $x$.

The calculation of $B(T)$ was performed by a Monte Carlo approach, referred to as path-integral Monte Carlo (PIMC), using a parallel implementation~\cite{parallel_vegas} of the
\mbox{VEGAS}~\cite{Lepage78} integration routine. We used 32 independent samples to evaluate the $\langle
\cdots \rangle$ average and $10^6$ integration steps, cutting off the interaction after
$200$~nm. The second virial coefficient for N${}_2$ was evaluated using at least 64 independent calculations
for $T > 1000$~K (up to 3000~K for N$_2$ and 2000~K for O$_2$), at least $96$ independent calculations for $60 \leq T / \mathrm{K} \leq 1000$, and
then progressively reducing the independent calculations to 10 at $T=10$~K.
In the case of O${}_2$, we used the same number of independent samples, integration steps, and
cutoff. The number of independent calculations for each spin state was at least $64$ for $T \geq
150$~K, progressively increased to an average of $96$ down to $T = 80$~K and then progressively
reduced to $10$ at lower temperatures.


\section{Results}\label{sec:results}

The values of $B$ calculated as described in Sec.~\ref{sec:calc} are given in the \textcolor{blue}{Supplementary Material} for both nitrogen and oxygen, along with the statistical uncertainty of the PIMC integration.
For convenience in use and to facilitate comparison with experimental data, these values were fitted to smooth functions as described in Sec.~\ref{sec:recc}.

\subsection{Nitrogen}\label{sec:results_N2}

\begin{figure*}
\includegraphics{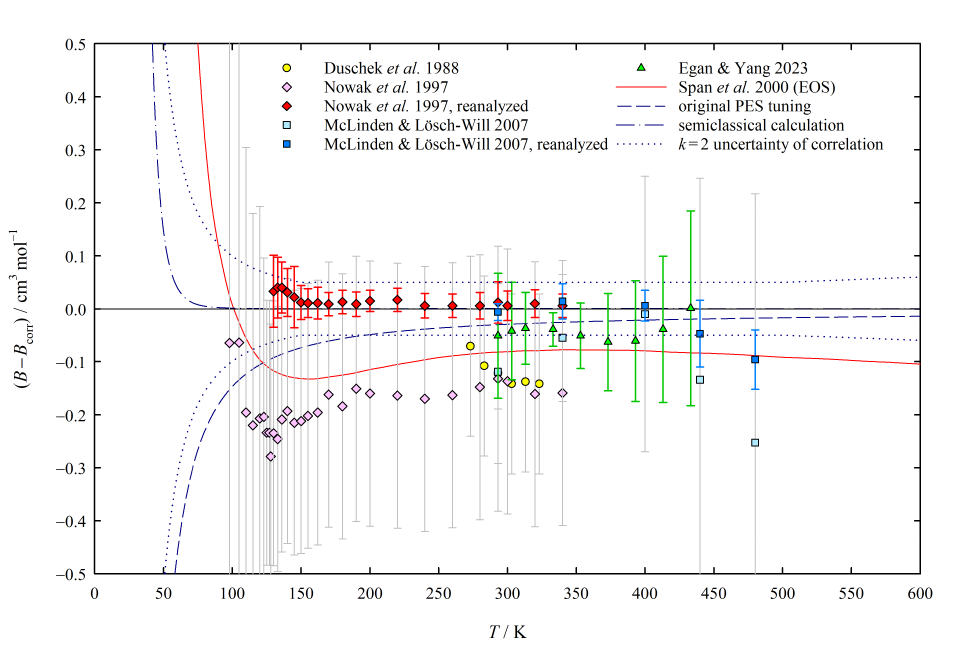}
\caption{Deviations of selected experimental data for the second virial coefficient $B$ of nitrogen,\cite{Duschek_1988,Nowak_1997,McLinden_2007,Egan_2023} including those derived from the reanalyses of this work, from a correlation (see Sec.~\ref{sec:recc}) of $B(T)$ calculated with the retuned PES by the PIMC approach. Also plotted are values of $B$ derived from the reference equation of state (EOS) for nitrogen of Span \textit{et al.},\cite{Span_2000} values obtained with the original tuning of the nitrogen pair PES, and values from the revised tuning of this PES computed by the semiclassical QFH approach. The dotted lines indicate the estimated expanded ($k=2$) uncertainty of the correlation.}
\label{fig:B_N2}
\end{figure*}

Figure~\ref{fig:B_N2}
shows the deviations of the experimental $B$ values for nitrogen of Duschek \textit{et al.},\cite{Duschek_1988} Nowak \textit{et al.}\cite{Nowak_1997} (original and reanalyzed), McLinden and L\"{o}sch-Will\cite{McLinden_2007} (original and reanalyzed), and Egan and Yang\cite{Egan_2023} from a correlation (see Sec.~\ref{sec:recc}) of the values calculated by the PIMC approach with the retuned PES. The calculated values agree with all $B$ values derived from the reanalyses of the measurements of Nowak \textit{et al.}\ and McLinden and L\"{o}sch-Will, except for the 480\;K value of McLinden and L\"{o}sch-Will, within the expanded ($k=2$) experimental uncertainties. The $B$ values of Egan and Yang have on average higher uncertainties and were therefore less strongly weighted in the tuning of the PES. Nevertheless, all but one value of Egan and Yang are described within their expanded ($k=2$) uncertainties. It is interesting that most of the $B$ values derived by Duschek \textit{et al.}\ and originally by Nowak \textit{et al.}\ and McLinden and L\"{o}sch-Will are significantly more negative than the new reference values. This is the result of a significant overestimation of the third virial coefficient $C$ by these groups due to the strong negative correlation between $B$ and $C$ when obtained simultaneously from fits to isothermal $p(\rho)$ data. The overestimation of $C$, on the other hand, is caused by its strong negative correlation with the positive fourth virial coefficient $D$, whose values were unknown at the time and therefore set to zero.

The small systematic offset of the values of $B$ from Egan and Yang\cite{Egan_2023} might result from the derivation of $B$ from their refractivity data.
The quantity more directly resulting from their experiments, measured with high accuracy, was the second-order coefficient in an expansion of the refractivity, denoted as $\mathcal{B}$, which is related to both density and refractivity virial coefficients by
\begin{equation}
 \mathcal{B} = \frac{2B}{3A_\mathrm{R}} - \frac{1}{6} - \frac{2B_\mathrm{R}}{3A_\mathrm{R}^2} ,
 \end{equation} 
where $A_\mathrm{R}$ and $B_\mathrm{R}$ are respectively the first and second refractivity virial coefficients. The single-molecule quantity $A_\mathrm{R}$ is known very accurately from experiment,\cite{Egan_2023} but $B_\mathrm{R}$ is difficult to measure independently and so contributes some uncertainty to the derived $B$.
While their paper does not correctly describe the source of the $B_\mathrm{R}$, Egan and Yang used $B_\mathrm{R}(T)$ based entirely on the work of Hohm,\cite{Hohm_1993,Egan_PC} who combined the experimental value of Burns \emph{et al.}\cite{Burns_1986} at 298.2\;K with a theoretically calculated temperature dependence.
If $B_\mathrm{R}$ were about 20\% higher than the values assumed (consistent with the experimental datum of Achtermann \textit{et al.}\cite{Achtermann_1991}), agreement with our recommendation would be much better.
It may be possible to resolve this issue with theory; work is underway\cite{MQB_web} to compute $B_\mathrm{R}$ based on \textit{ab initio} surfaces for the two-body polarizability and dispersion.

Figure~\ref{fig:B_N2} also shows the deviations of the $B$ values obtained from the reference equation of state (EOS) of Span \textit{et al.},\cite{Span_2000} the values resulting from the original tuning of the pair PES, and the values obtained with the revised tuning by the semiclassical QFH approach. The EOS in the gas phase is primarily based on the data of Nowak \textit{et al.}\cite{Nowak_1997} That it gives somewhat higher $B$ values than the virial analysis performed by Nowak \textit{et al.}\ could be due to the fact that the EOS implicitly accounts for fourth and higher virial coefficients and thus is less affected by the shortcomings of the original truncated virial analysis.
The original tuning of the PES performed by Hellmann\cite{Hellmann_2013} yields more negative values of $B$ at all temperatures. It was a compromise between the original Nowak \textit{et al.}\ values and a few other data sets that are not shown in Fig.~\ref{fig:B_N2} but are given in Ref.~\citenum{Hellmann_2013}. The semiclassical QFH approximation yields results essentially identical to the full PIMC calculation over most of the investigated temperature range, which strongly validates the correctness of these calculations. The QFH values begin to differ significantly at temperatures below about 60\;K, but even at 20\;K the relative difference is only about 1\,\%.

\begin{figure}
\includegraphics{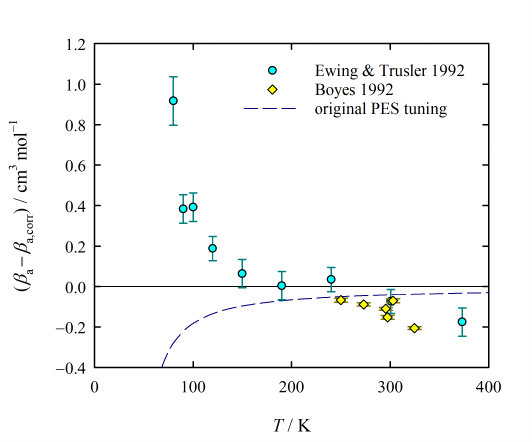}
\caption{Deviations of the available experimental data for the second acoustic virial coefficient $\beta_\mathrm{a}$ of nitrogen\cite{Ewing_1992,Boyes_1992} and of $\beta_\mathrm{a}(T)$ derived from a correlation of the $B$ values obtained with the original tuning of the nitrogen pair PES from $\beta_\mathrm{a}(T)$ derived from a correlation (see Sec.~\ref{sec:recc}) of the $B(T)$ obtained with the retuned PES by the PIMC approach.}
\label{fig:beta_a_N2}
\end{figure}

In Fig.~\ref{fig:beta_a_N2},
we show the deviations of the experimental data of Ewing and Trusler\cite{Ewing_1992} and Boyes\cite{Boyes_1992} for the second acoustic virial coefficient $\beta_\mathrm{a}$ from the values obtained from our $B(T)$ correlation using the exact relation
\begin{equation}
\label{eq:beta_a}
\beta_\mathrm{a}(T) = 2 B(T)+2(\gamma^\circ - 1) T \frac{\mathrm{d}B(T)}{\mathrm{d}T} + \frac{(\gamma^\circ - 1)^2}{\gamma^\circ}T^2 \frac{\mathrm{d}^2B(T)}{\mathrm{d}T^2},
\end{equation}
where $\gamma^\circ=c_p^\circ/c_V^\circ$ is the ratio of the isobaric and isochoric ideal-gas heat capacities, which is very close to $7/5$ for nitrogen in the temperature range of the experimental data.\cite{Gamache_2023}
Also shown are values derived from a correlation of the values for the original tuning of the PES, which gives an indication of the sensitivity of $\beta_\mathrm{a}(T)$ to small changes in the PES. 
The comparison with the two data sets, which are the only ones available for $\beta_\mathrm{a}$, is somewhat disappointing.
The low-temperature systematic deviations of Ewing and Trusler's values and the deviation of their datum at the highest temperature cannot reasonably be explained by a deficiency of our correlation. In the case of the data of Boyes, it is clear that their uncertainties are severely underestimated, as the three data points around 300\;K are mutually inconsistent. Therefore, we see a need for new state-of-the-art measurements of $\beta_\mathrm{a}$ for nitrogen (and oxygen, for which such data do not exist at all) to improve our knowledge of $B$ and $\beta_\mathrm{a}$ beyond the level achieved in this work.

\subsection{Oxygen}\label{sec:results_O2}

\begin{figure}
\includegraphics{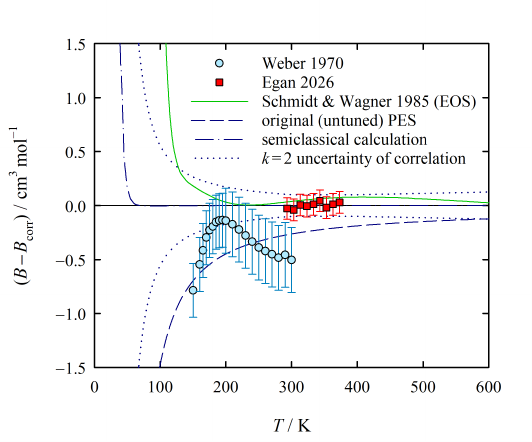}
\caption{Deviations of the experimental data of Weber\cite{Weber_1970} and Egan\cite{Egan_2026} for the second virial coefficient $B$ of oxygen from a correlation (see Sec.~\ref{sec:recc}) of the $B(T)$ calculated with the tuned PES by the PIMC approach. Also plotted are values of $B$ derived from the reference EOS for oxygen of Schmidt and Wagner,\cite{Schmidt_1985} values obtained with the three original (untuned) oxygen pair PES, and values calculated from the tuned versions of these PES by the semiclassical QFH approach. The dotted lines indicate the estimated expanded ($k=2$) uncertainty of the correlation.}
\label{fig:B_O2}
\end{figure}

Figure~\ref{fig:B_O2}
shows the deviations of the experimental values of $B$ for oxygen of Weber\cite{Weber_1970} and of the very recent data of Egan\cite{Egan_2026} from a correlation (see Sec.~\ref{sec:recc}) of the values calculated using the PIMC approach with the three PES tuned to the data of Egan. The agreement with the values of Egan, which have an expanded ($k=2$) uncertainty of 0.1\;cm$^3$\,mol$^{-1}$, is within $\pm0.05$\;cm$^3$\,mol$^{-1}$. For the data of Weber, which were previously regarded as the most accurate available for oxygen, the agreement is only partly within the claimed experimental uncertainties. The values obtained with the three original (untuned) PES, which are also shown in the figure, agree better with the data of Weber but are not consistent with the data of Egan within the latter's expanded uncertainty. For a comparison with further experimental data sets, we refer the reader to the paper of Hellmann.\cite{Hellmann_2023}

Figure~\ref{fig:B_O2} also compares with values of $B(T)$ derived from the reference EOS of Schmidt and Wagner\cite{Schmidt_1985} and those obtained from the tuned PES by the semiclassical QFH approach. The agreement with the EOS values above about 120\;K is remarkably good despite the fact that the data of Egan were not available when the EOS was developed. At lower temperatures, the deviations increase dramatically, and, as shown by Hellmann,\cite{Hellmann_2023} the EOS values eventually become completely unphysical. The $B$ values obtained with the QFH approximation again substantiate the PIMC calculations and begin to deviate significantly from the PIMC values only below about 50\;K.


\section{Estimates of Uncertainty}\label{sec:unc}

The uncertainties of our recommended $B(T)$ are based on the uncertainty of the experimental data used to tune the potential (in the temperature range of those data) and on our estimate of the uncertainty at other temperatures due to the uncertainties of the tuned potentials.
In principle, there are also contributions from the statistical uncertainty of the PIMC calculations and from the deviations from the calculated data of our $B(T)$ fits described in Sec.~\ref{sec:recc}.
However, both of these contributions are relatively negligible and will therefore not be considered in the following.

For nitrogen $B(T)$, we estimate the expanded uncertainty at the $k=2$ level to be 0.05\;cm$^3$\,mol$^{-1}$ in the temperature range 150--500\;K.
This is based on the level of agreement with the experimental sources used to tune the potentials (see Sec.~\ref{sec:potentials_N2}) and the uncertainties of the experimental values. 
At temperatures below this range, our uncertainty estimates are based on the observation that the possible error in the PES is constrained by the 0.05\;cm$^3$\,mol$^{-1}$ expanded uncertainty at 150~K.
We therefore estimate the expanded uncertainty at low temperatures based on the difference between values of $B$ obtained with the original tuning and the revised tuning (at the QFH level, which is accurate enough to compute differences), rescaled by a constant factor such that at 150\;K a difference of 0.05\;cm$^3$\,mol$^{-1}$ is obtained.
Above 500~K, the difference in $B$ between the original tuned PES and the retuned PES becomes small, so that in our opinion the dominant uncertainty becomes the effect of neglected molecular flexibility, which will grow (slowly) with temperature. 
Therefore, our estimate for the expanded uncertainty increases linearly from 0.05\;cm$^3$\,mol$^{-1}$ at 500\;K to 0.3\;cm$^3$\,mol$^{-1}$ at 3000\;K. This is a conservative estimate, because the effect of flexibility on $B(T)$ was found to be smaller than this for the less rigid H$_2$ molecule.\cite{Garberoglio2014}

The procedure for estimating the expanded ($k=2$) uncertainty of $B(T)$ for oxygen is similar.
In the range 300--400\;K, our estimate is 0.1\;cm$^3$\,mol$^{-1}$ based on agreement with the data of Egan\cite{Egan_2026} and the uncertainty of those data. 
Below 300 K, we use the difference between values of $B$ obtained with and without tuning, rescaled such that at 300\;K a difference of 0.1\;cm$^3$\,mol$^{-1}$ is obtained. 
Above 400\;K, our estimate (again based on increasing effects of flexibility at high temperatures) is a linear increase from 0.1\;cm$^3$\,mol$^{-1}$ at 400\;K to 0.3\;cm$^3$\,mol$^{-1}$ at 2000\;K.


\section{Representation of Recommended Values}\label{sec:recc}

We provide recommended values for the second virial coefficients of nitrogen and oxygen as functions of temperature through correlations of the form
\begin{equation}
\label{eq:Bfit}
\frac{B(T)}{\mathrm{cm}^3\,\mathrm{mol}^{-1}} = \sum_{i=1}^{i_\mathrm{max}} a_{B,i}\left(T^*\right)^{b_{B,i}},
\end{equation}
where $i_\mathrm{max}=10$ for N$_2$ and $i_\mathrm{max}=11$ for O$_2$, $T^*=T/(100\;\mathrm{K})$, and $a_{B,i}$ and $b_{B,i}$ are fit parameters, with the $b_{B,i}$ constrained to be quadratic functions of the index $i$ 
(i.e., $b_{B,i} = c_{1} + c_{2}\,i + c_{3}\,i^2$) in order to maintain independence of the terms by keeping the exponents $b_{B,i}$ well separated.
The individual $B$ values were weighted in the fits by the squared inverses of the statistical uncertainties resulting from the PIMC calculations. The number of summands $i_\mathrm{max}$ in Eq.~\eqref{eq:Bfit} was chosen such that it is just large enough for the fit residuals to be of similar magnitude to the statistical uncertainties and to scatter randomly. The parameters $a_{B,i}$ and $b_{B,i}$ are listed in Table~\ref{tab:correlations}.
\begin{table*}
\footnotesize
\caption{\label{tab:correlations} Parameters of Eqs.~\eqref{eq:Bfit} and \eqref{eq:uBfit}.} 
\begin{ruledtabular}
\begin{tabular}{lllll}
\multicolumn{1}{c}{$i$} &
\multicolumn{1}{c}{$a_{B,i}$} &
\multicolumn{1}{c}{$b_{B,i}$} &
\multicolumn{1}{c}{$a_{U(B),i}$} &
\multicolumn{1}{c}{$b_{U(B),i}$}
\\[0.5ex]
\hline
\\[-1.5ex]
\multicolumn{5}{c}{Nitrogen} \\[1ex]
1  & $\phantom{-}0.377901252\times 10^{3}$ & $-0.272248910             $ & $0.782831\times 10^{-1}$ & $-0.133461\times 10^{1}$ \\
2  & $-0.554851973\times 10^{2}          $ & $-0.103437798             $ & $0.212489\times 10^{-1}$ & $-0.386759\times 10^{1}$ \\
3  & $-0.315831429\times 10^{3}          $ & $-0.414337589             $ & $0.869303\times 10^{-4}$ & $-0.774783\times 10^{1}$ \\
4  & $-0.120574085\times 10^{3}          $ & $-0.120494828\times 10^{1}$ & $                      $ & $$ \\
5  & $-0.406097668\times 10^{2}          $ & $-0.247526988\times 10^{1}$ & $                      $ & $$ \\
6  & $-0.542293842\times 10^{1}          $ & $-0.422530239\times 10^{1}$ & $                      $ & $$ \\
7  & $-0.165143509                       $ & $-0.645504579\times 10^{1}$ & $                      $ & $$ \\
8  & $-0.911990703\times 10^{-3}         $ & $-0.916450010\times 10^{1}$ & $                      $ & $$ \\
9  & $-0.636887692\times 10^{-6}         $ & $-0.123536653\times 10^{2}$ & $                      $ & $$ \\
10 & $-0.735438549\times 10^{-10}        $ & $-0.160225414\times 10^{2}$ & $                      $ & $$ \\[1ex]
\multicolumn{5}{c}{Oxygen} \\[1ex]
1  & $\phantom{-}0.200532304\times 10^{3}$ & $-0.278782772             $ & $0.378462              $ & $-0.123132\times 10^{1}$ \\
2  & $-0.403578707\times 10^{2}          $ & $-0.154921077             $ & $0.175960              $ & $-0.400677\times 10^{1}$ \\
3  & $-0.213340129\times 10^{3}          $ & $-0.601115805             $ & $0.358783\times 10^{-3}$ & $-0.908617\times 10^{1}$ \\
4  & $-0.105898022\times 10^{3}          $ & $-0.161736696\times 10^{1}$ & $                      $ & $$ \\
5  & $-0.320604731\times 10^{2}          $ & $-0.320367453\times 10^{1}$ & $                      $ & $$ \\
6  & $-0.276935482\times 10^{1}          $ & $-0.536003853\times 10^{1}$ & $                      $ & $$ \\
7  & $-0.454509753\times 10^{-1}         $ & $-0.808645895\times 10^{1}$ & $                      $ & $$ \\
8  & $-0.135898524\times 10^{-3}         $ & $-0.113829358\times 10^{2}$ & $                      $ & $$ \\
9  & $-0.629692003\times 10^{-7}         $ & $-0.152494691\times 10^{2}$ & $                      $ & $$ \\
10 & $-0.404246166\times 10^{-11}        $ & $-0.196860588\times 10^{2}$ & $                      $ & $$ \\
11 & $-0.282544248\times 10^{-16}        $ & $-0.246927049\times 10^{2}$ & $                      $ & $$ \\[1ex]
\end{tabular}
\end{ruledtabular}
\end{table*}
We note that if Eq.~(\ref{eq:Bfit}) is implemented correctly, it yields $B=-15960.532$\;cm$^3$\,mol$^{-1}$ for N$_2$ at 20\;K, $B=-4.4726027$\;cm$^3$\,mol$^{-1}$ for N$_2$ at 300\;K, $B=-58543.392$\;cm$^3$\,mol$^{-1}$ for O$_2$ at 20\;K, and $B=-15.506385$\;cm$^3$\,mol$^{-1}$ for O$_2$ at 300\;K.

Even though the parameters were fitted to all PIMC values for $B$ down to 10\;K, we fully endorse the correlations only from 20\;K upwards. Our reservation stems from the fact that the magnitude of $B$ and its relative statistical uncertainty increase so rapidly toward the lowest temperatures that the correlations may no longer provide reliable interpolation there. However, Eq.~(\ref{eq:Bfit}) still behaves in a physically correct manner even below 10\;K because the coefficients $a_{B,2}$ to $a_{B,i_\mathrm{max}}$ are all negative. Above the highest considered temperatures, i.e., 3000\;K for N$_2$ and 2000\;K for O$_2$, the extrapolation is also physically reasonable, with $B(T)$ remaining bounded due to all exponents $b_{B,i}$ being negative.

We fitted the expanded ($k=2$) uncertainties of $B$, $U(B)$, for temperatures below 150\;K for N$_2$ and below 300\;K for O$_2$ using the same functional form as above but with only three terms,
\begin{equation}
\label{eq:uBfit}
\frac{U\left[B(T)\right]}{\mathrm{cm}^3\,\mathrm{mol}^{-1}} = \sum_{i=1}^{3} a_{U(B),i}\left(T^*\right)^{b_{U(B),i}},
\end{equation}
where the parameters $a_{U(B),i}$ and $b_{U(B),i}$ are given in Table~\ref{tab:correlations}. As stated in Sec.~\ref{sec:unc}, we have $U(B)=0.05$\;cm$^3$\,mol$^{-1}$ for nitrogen from 150--500\;K and $U(B)=0.1$\;cm$^3$\,mol$^{-1}$ for oxygen from 300--400\;K. The linearly increasing expanded ($k=2$) uncertainty estimates above these temperature ranges can be written for nitrogen as
\begin{equation}
\label{eq:uBhigh_1}
\frac{U\left[B(T)\right]}{\mathrm{cm}^3\,\mathrm{mol}^{-1}} = 0.01\,T^*,
\end{equation}
and for oxygen as
\begin{equation}
\label{eq:uBhigh_2}
\frac{U\left[B(T)\right]}{\mathrm{cm}^3\,\mathrm{mol}^{-1}} = 0.05 + 0.0125\,T^*.
\end{equation}


\section{Conclusions}\label{sec:concl}
\textit{Ab initio} pair potentials for molecular nitrogen and oxygen have been tuned based on selected high-accuracy experimental data.
The tuning took advantage of recent data based on precise refractometry for both nitrogen\cite{Egan_2023} and oxygen.\cite{Egan_2026}
For nitrogen, the most important contributions to the tuning came from values of $B$ derived from state-of-the-art gas density measurements\cite{Nowak_1997,McLinden_2007} with the aid of \textit{ab initio} virial coefficients $D$, $E$, and $F$ calculated with a new three-body potential for N$_2$ as described in Sec.~\ref{sec:potentials_N2}.
In both cases, the tuning perturbed the potentials by amounts smaller than the uncertainties of the original \textit{ab initio} PES.

The tuned potentials were used to calculate the second virial coefficient $B(T)$ for both species using the PIMC approach to fully account for nuclear quantum effects; the temperature range of the calculations was 10--3000~K for N$_2$ and 10--2000~K for O$_2$.
The representations for $B(T)$ and its uncertainty in Sec.~\ref{sec:recc} are recommended for use from 20--3000~K for N$_2$ and 20--2000~K for O$_2$ and provide state-of-the-art information for this important thermodynamic quantity.

The recommended $B(T)$ given here will reduce the uncertainty of thermodynamic properties for these gases in applications such as those mentioned in Sec.~\ref{sec:intro}, such as flow metering and pressure metrology.
In addition, they provide low-density boundary conditions that can be used for future improvement of the reference equations of state for these fluids; the current reference EOS for nitrogen\cite{Span_2000} is over 25 years old while that for oxygen\cite{Schmidt_1985} is over 40 years old.

In addition to the second virial coefficient, the original papers reporting PES for these systems\cite{Hellmann_2013,Hellmann_2023} reported calculations for the low-density limits of the viscosity and thermal conductivity.
In principle, those quantities could be recalculated with the tuned potentials reported in this work.
However, the difference due to the tuning would be much smaller than the uncertainty of the transport calculations.
Therefore, the viscosity and thermal conductivity from the earlier work can be used with no changes.

The recommended $B(T)$ developed for N$_2$ and O$_2$ in this work will form a foundation for a revised recommendation for the second virial coefficient of (dry) air, $B_\mathrm{aa}(T)$.
The other important contributor to this quantity is the cross second virial coefficient between N$_2$ and O$_2$. 
A project is underway to derive a state-of-the-art PES for this pair, and the resulting surface can be tuned to data of Egan,\cite{Egan_2026} whose refractometry measurements resulted in values of $B(T)$ with low uncertainties for both binary N$_2$/O$_2$ mixtures and natural air.
Calculation of $B_\mathrm{aa}$ will also require accounting for the roughly 1\,\% of argon in air. $B(T)$ for pure argon is known with good accuracy from a recent pair potential\cite{Lang_2024} but makes a tiny contribution, and the small contributions of Ar/N$_2$ and Ar/O$_2$ interactions should be obtainable with sufficient accuracy from PES that already exist or that can be developed with relatively little effort.

Finally, a recommended $B_\mathrm{aa}(T)$ for dry air is a necessary step toward the description of humid air, which is needed in many applications.
In addition to $B_\mathrm{aa}$, describing humid air requires the pure water virial coefficient $B_\mathrm{ww}$ and the water--air cross coefficient $B_\mathrm{aw}$.
Harvey and Lemmon gave a formulation for $B_\mathrm{ww}(T)$ in 2004,\cite{Harvey_2004} but it was only fitted to data above 313~K (and the data had large uncertainties below about 350~K), so there is room to improve its low-temperature behavior with state-of-the-art \textit{ab initio} pair potentials.
$B_\mathrm{aw}(T)$ is known with good accuracy from the work of Hellmann,\cite{Hellmann_2020} who constructed the function based on second virial coefficients calculated from high-accuracy pair potentials for H$_2$O--N$_2$, H$_2$O--O$_2$, H$_2$O--Ar, and H$_2$O--CO$_2$.

\section{Supplementary Material}

See the \textcolor{blue}{Supplementary Material} for Fortran~90 codes of the tuned nitrogen and oxygen pair potentials of this work and for the discrete $B(T)$ points calculated with the PIMC approach for nitrogen and oxygen.

\section*{Acknowledgments}
The work of RH and GG was supported by the project ``Metrology for quantum-based traceability of the pascal'' (22IEM04 MQB-Pascal), which has received funding from the European Partnership on Metrology, co-financed from the European Union's Horizon Europe Research and Innovation Programme and by the Participating States. RH used computational resources (HPC cluster HSUper) provided by the project hpc.bw, funded by dtec.bw -- Digitalization and Technology Research Center of the Bundeswehr. dtec.bw is funded by the European Union -- NextGenerationEU.
NIST work was funded solely by the United States Government.
GG thanks the University of Trento for a generous allocation of CPU time on their HPC cluster.
We thank Patrick Egan for making the data from Ref.~\citenum{Egan_2026} available prior to publication and for several helpful suggestions.

\section{Author Declarations}
\subsection{Conflict of interest}
The authors have no conflicts to declare.

\section{Data Availability}
Most of the data that support the findings of this study are available within the article and its supplementary material.
The nonadditive three-body PES for N$_2$ used for calculating higher virial coefficients is available upon reasonable request to RH. 

\bibliography{N2O2}

\end{document}